\documentclass[11pt,reqno]{amsart}
\usepackage{amsfonts}
\usepackage{mathrsfs}
\usepackage{amsmath}
\usepackage{amssymb,amsfonts}

\numberwithin{equation}{section}

\newfont{\bb}{msbm10 at 12pt}

\def\R{\hbox{\bb R}}

\def\P{\mathcal P}

\newcommand{\bal}{\begin{aligned}}      \newcommand{\eal}{\end{aligned}}
\newcommand{\ba}{\begin{array}}      \newcommand{\ea}{\end{array}}
\newcommand{\bc}{\begin{center}}     \newcommand{\ec}{\end{center}}
\newcommand{\be}{\begin{enumerate}}  \newcommand{\ee}{\end{enumerate}}
\newcommand{\beq}{\begin{eqnarray}}  \newcommand{\eeq}{\end{eqnarray}}
\newcommand{\beQ}{\begin{eqnarray*}} \newcommand{\eeQ}{\end{eqnarray*}}
\newcommand{\bi}{\begin{itemize}}    \newcommand{\ei}{\end{itemize}}
\newcommand{\bt}{\begin{tabular}}    \newcommand{\et}{\end{tabular}}
\newcommand{\bdm}{\begin{displaymath}} \newcommand{\edm}{\end{displaymath}}

\begin{document}

\title[]{Energy-momentum in general relativity}

\author{Xiaoning Wu, Xiao Zhang}
\address[]{Institute of Mathematics, Academy of Mathematics and
Systems Science, Chinese Academy of Sciences, Beijing 100190, PR China}
\email{wuxn@amss.ac.cn,xzhang@amss.ac.cn}

\date{}

\begin{abstract}
We briefly review of the definitions of the total energy, the total linear momentum and the angular momentum of gravitational field when the cosmological constant is zero. In particular, we show pseudo-tensor's definition of the energy and the momentum given by Prof. Duan in 1963 agree with the ADM total energy-momentum and the Bondi energy-momentum at spatial and null infinity respectively. We also review the relevant energy-momentum inequalities. Finally, we provide a short review of the positive energy theorem and the peeling property of the Newmann-Penrose quantities when the cosmological constant is positive.

\end{abstract}

\maketitle \pagenumbering{arabic}

\section{Energy-momentum of gravitational systems}

\subsection{The total conserved quantities for matter fields}

It is well known that conserved quantities, such as energy and momentum, are very important both in physics and in mathematics. For ordinary matter fields, people usually use the energy-momentum tensor $T_{ab}$ to describe the distribution of energy and momentum. Such tensor comes from the variation of Lagrangian function of the matter field. For simplicity, we only consider Lagrangian which contains the physical field $\phi$ and its first order derivative, i.e. $L_m=L_m(\phi,\nabla\phi,g_{ab})$, $\phi$ is a $(p,q)$ type tensor field. The space-time metric $g_{ab}$ in the Lagrangian is only served as a geometric back ground, which means the physical field $\phi$ is living on a fixed space-time and the back reaction of the matter field to the space-time is neglected. Since the Lagrangian is a scalar function on space-time, it should be natural to require that it is diffeomorphism invariant, i.e. $L_m(\psi^*_t[g_{ab}],\psi^*_t[\phi],\psi^*_t[\nabla\phi])=\psi^*_t[L_m(g_{ab},\phi,\nabla\phi)]$ holds for any diffeomorphism map $\psi^*_t$. Given the Lagrangian $L_m$, the energy-momentum tensor can be calculated as following\cite{Sz09}:
\begin{eqnarray}
T_{ab}(\phi)&:=&\frac{2}{\sqrt{|g|}}\frac{\delta L_m}{\delta g^{ab}}\nonumber\\
&=&2\frac{\partial L_m}{\partial g^{ab}}-g_{ab}L_m\nonumber\\
&&+\frac{1}{2}\nabla^e(\sigma_{abe}
+\sigma_{bae}-\sigma_{aeb}-\sigma_{bea}-\sigma_{eab}-\sigma_{eba}).\label{EMT}
\end{eqnarray}
where $\sigma_{eab}$ is called canonical spin tensor and is defined as
\begin{eqnarray}
\sigma^{ea}_b:=(-1)\left(\frac{\partial L_m}{\partial\nabla_e\phi^{c\cdots}_{d\cdots}}\Delta^{ac\cdots g\cdots}_{bd\cdots h\cdots}\phi^{h\cdots}_{g\cdots}\right).
\end{eqnarray}
This term appears because of the deformation of the covariant derivative $\nabla$. $\Delta^{ac\cdots g\cdots}_{bd\cdots h\cdots}$ is the $(p+q+1,p+q+1)$ type invariant tensor, which is built from the Kronecker deltas.
Some simple calculation will show that the diffeomorphism invariance of $L_m$ ensures that $T_{ab}$ is divergence-free. If $O$ is a physical observer, he can measure the energy density $\rho$ and energy flux $j^a$ at each point of space-time, which can expressed as
\begin{eqnarray}
&&\rho:=T(e_0,e_0),\nonumber\\
&&j_i:=T(e_0,e_i),\quad i=1,2,3,
\end{eqnarray}
where $e_0$ is the 4-velocity of this observer and ${e_i}$ are normal basis chosen by him. If $\Sigma$ is a Cauchy surface, one may naively think the total energy of field $\phi$ on $\Sigma$ should be
\begin{eqnarray}
E=\int_{\Sigma}\rho.\label{naive}
\end{eqnarray}
Unfortunately, such simple idea is NOT correct. A simple observation is above defined total energy $E$ may not be conserved although the energy-momentum tensor is divergence-free. One can understand such problem in Minkowski case. If the background space-time is Minkowski and $O$ are inertial observers, it is easy to show that the energy defined by (\ref{naive}) is conserved. If $O$ are not inertial observer, the conservativeness of $E$ is broken. The key difference between the two kinds of observers is that the 4-velocity of initial observers is a Killing vector field. In fact, it is quite difficult to define a conserved total energy for general observers. For stationary space-time, a suitable definition of the total energy on $\Sigma$ is
\begin{eqnarray}
E=\int_{\Sigma}T(t, n)dV,
\end{eqnarray}
where $t^a$ is the time-like Killing vector and $n$ is the normal vector of $\Sigma$. Using the property of Killing vector field, it is easy to show that such energy is conserved. For non-stationary case, such conserved total energy does not exist. A physical explanation for this phenomena is that the non-stationary back ground gravitational filed will inject energy to the matter filed which makes the total energy of $\phi$ non-constant.

Similarly, one can also define the linear momentum $P_i$ measured by an observer $O$ as
\begin{eqnarray}
P_i:=T(e_0, e_i).
\end{eqnarray}
If $\Sigma$ is still Cauchy surface, a space-like Killing vector $k$ will give a conserved quantity $P_k$ as
\begin{eqnarray}
P_k:=\int_{\Sigma}T(k,n)dV.
\end{eqnarray}

We notice that existence of a conserved quantity is always associated with the existence of a Killing vector. This is nothing new but result of the well-known Noether theorem. This result also has a more direct way to see. If $\xi$ is a Killing vector, it will generate a diffeomorphism map $\psi^*_t$ which satisfies $\psi^*_t[g_{ab}]=g_{ab}$ and $[\psi^*_t,\nabla]=0$. The diffeomorphism invariance imply
\begin{eqnarray}
\psi^*_t[L_m]=L_m(g_{ab},\psi^*_t[\phi],\nabla\psi^*_t[\phi]).
\end{eqnarray}
Taking the $t$ derivative, one gets
\begin{eqnarray}
&&\xi^{\mu}\nabla_{\mu}L_m=\left(\frac{\partial L_m}{\partial\phi}-\nabla_a\frac{\partial L_m}{\partial\nabla_a\phi}\right)L_{\xi}\phi+\nabla_a\left(\frac{\partial L_m}{\partial\nabla_a\phi}L_{\xi}\phi\right),\nonumber\\
&\Rightarrow& \nabla_a\left[\left(\frac{\partial L_m}{\partial\nabla_a\phi}L_{\xi}\phi\right)-\xi^aL_m\right]=\left(\frac{\partial L_m}{\partial\phi}-\nabla_a\frac{\partial L_m}{\partial\nabla_a\phi}\right)L_{\xi}\phi.
\end{eqnarray}
If $\phi$ is a solution of Euler-Lagrangian equation, $J^a[\xi,\phi]:=\left(\frac{\partial L_m}{\partial\nabla_a\phi}L_{\xi}\phi\right)-\xi^aL_m$ is a conserver current. On any Cauchy surface, there is a conserved quantity $Q[\xi,\phi]$ as
\begin{eqnarray}
Q[\xi,\phi]:=\int_{\Sigma}J\cdot d\Sigma.
\end{eqnarray}
One can prove\cite{Sz09} that $E=Q[t,\phi]$ and $P_k=Q[k,\phi]$. Furthermore, the diffeomorphism invariance of $L_m$ also implies another equation which is called Noether identity. Consider a smooth vector field $K^a$, the diffeomorphism invariance of the action implies\cite{Sz09,IW94,IW95}
\begin{eqnarray}
&&\left(\frac{\partial L_m}{\partial\phi}-\nabla_a\frac{\partial L_m}{\partial\nabla_a\phi}\right) L_K\phi+\frac{1}{2}T_{ab}L_Kg^{ab}+\nabla_eC^e=0,\label{canoEMT}\\
&&C^e[K,\phi]=\theta^{ec}K_c+\left(\sigma^{e[ab]}+\sigma^{a[be]}+\sigma^{b[ae]}\right)\nabla_aK_b,\nonumber\\
&&\theta^{ab}=-L_mg^{ab}+\frac{\partial L_m}{\partial\nabla_a\phi}\nabla_b\phi.\nonumber
\end{eqnarray}
where $\theta^{ab}$ is called {\it canonical energy-momentum tensor} and $C^a[K]$ is called {\it canonical Noether current}. The difference between $\theta_{ab}$ and $T_{ab}$ is some divergence term. Unlike $T_{ab}$, $\theta_{ab}$ contains some ``unphysical" information. Different choice of gauge, coordinates or boundary condition may give different $\theta_{ab}$. This is also the reason why people can construct many different type energy definition for gravitational field \cite{Sz09,CNT15}.
Starting from this equation and using the fact that $\phi$ is a solution and $K$ is Killing vector, one can get the conserved current $J^a$ again.

\subsection{The total conserved quantities for gravity}
In last section, we discussed the total energy and momentum of a matter field. The Noether's principle is a nice tool to define conserved quantities. For gravity case, things become more complex. For matter case, the space-time $(M,g_{ab})$ is only served as a background and does not involve the dynamics. The non-dynamical background structure enables us to do variation for dynamical variables $\phi$ and for background $g_{ab}$ separately and corresponding results are the dynamical equation and the energy-momentum tensor. If we consider gravity, $\phi$ and $g_{ab}$ are all dynamical variable so we have no natural energy-momentum tensor for gravitational field. This fact also can be seen as the result of equivalent principle. Roughly speaking, Eq.(\ref{EMT}) shows that the energy-momentum tensor of matter field will depend on the first derivative of the field variables in quadratic form, but the existence of Riemann normal coordinates implies such term will always vanish. This fact implies that one need some ``additional structure" to introduce the conserved quantities for gravity.

The first ``additional structure" is Killing vector. As we have noticed in last section, the existence of Killing vector is helpful for finding conserved quantities. With the help of Killing vector, one can introduce conserved quantities following the idea of Noether theorem.

Consider the Einstein-Hilbert action
\begin{eqnarray}
I_{EH}=\frac{1}{16\pi}\int_MR\sqrt{-g}dx^4.
\end{eqnarray}
Obviously, $L_{EH}=R\sqrt{-g}$ is diffeomorphism invariant. Suppose there exists a Killing vector field $\xi$ on space-time, $\psi^*_t$ is the difeomorphism map generated by $\xi$,
\begin{eqnarray}
0&=&\xi^a\nabla_a(R\sqrt{-g})\nonumber\\
&=&[g^{ac}\delta R_{ab}+G_{ab}\delta g^{ab}]\sqrt{-g}\nonumber\\
&=&\nabla^a[\nabla^bL_{\xi}g_{ab}-g^{bc}\nabla_aL_{\xi}g_{bc}]\nonumber\\
&=&\nabla^a(R_{ab}\xi^b).
\end{eqnarray}
So we get a conserved current $J_a=R_{ab}\xi^b$. Using the Killing equation, such current can be rewritten as
\begin{eqnarray}
J&=&\nabla^b\nabla_a\xi_b-\nabla^b\nabla_b\xi_a\nonumber\\
&=&\nabla^b[\delta^e_a\delta^f_b-\delta^e_b\delta^f_a]\nabla_e\xi_f\nonumber\\
&=&\epsilon_{abcd}\nabla^b\epsilon^{cdef}\nabla_e\xi_f\nonumber\\
&=&*d*d\xi.
\end{eqnarray}
Using Stokes' theorem, the conserved charge Q is
\begin{eqnarray}
Q[\xi]=\frac{1}{8\pi}\int_{\Sigma}d*d\xi=\frac{1}{8\pi}\int_{\partial\Sigma}*d\xi.
\end{eqnarray}
This is the famous Komar integral which is found by Komar in 1959 \cite{Ko59}. If $\xi$ is a time-like Killing vector, $Q[\xi]$ is the famous Komar mass. If $\xi$ is axial-symmetric Killing vector, $Q[\xi]$ is Komar angular-momentum.

Second possible ``additional structure" is coordinates, or in other words replacing the the Levi-Civita covariant derivative by the ordinary partial derivative. In a fixed local coordinates, since the gravitational Lagrangian is diffeomorphism invariant, one can use Eq.(\ref{canoEMT}) to get the canonical energy-momentum tensor. Since such tensor depends on local coordinates, it is not covariant under the coordinate transformation. This is the famous pseudotensors method. In the early time of pseudotensors method, people get the pseudotensors just by rewritting the Einstein equation, for example see \cite{Ein15,LL62,Wein72}. It is found that the Einstein equation can be rewritten into following form under suitable choice of coordinates,
\begin{eqnarray}
16\pi(\sqrt{-g}T^a_b+t^a_b)=\partial_aU^{ac}_b,
\end{eqnarray}
where $T_{ab}$ is the energy momentum tensor of matter field, $U^{ac}_b$ is called {\it superpotential} and satisfies $U^{ac}_b=U^{[ac]}_b$. Using the symmetric property of $U^{ac}_b$, it is easy to see that
\begin{eqnarray}
\partial_a(\sqrt{-g}T^a_b+t^a_b)=0.
\end{eqnarray}
Since $T_{ab}$ is divergence-free, above equation implies
\begin{eqnarray}
\partial_at^a_b=0.
\end{eqnarray}
$t^a_b$ is called {\it the energy-momentum pseudotensor} of gravitational filed. Weinberg gave a more direct way to see why people call $t^a_b$ energy-momentum pseudotensor \cite{Wein72}. He introduces a "Minkowski-like" coordinates in asymptotic flat space-time and reexpresses the metric as $g_{ab}=\eta_{ab}+h_{ab}$. If $h_{ab}$ is small enough, the Einstein equation becomes
\begin{eqnarray}
R^{(1)}_{ab}-\frac{1}{2}R^{(1)}\eta_{ab}=8\pi(T_{ab}+t_{ab}),
\end{eqnarray}
where $R^{(1)}_{ab}-\frac{1}{2}R^{(1)}\eta_{ab}$ is the linear term of $h_{ab}$ in Einstein tensor. In this equation, it is quite clear that $t_{ab}$ is contributed by the curved geometry and can be seen as the energy-momentum pseudotensor. Since the energy-momentum pseudotensor is related with different coordinate choice, there are many different kinds of energy-momentum pseudotensor which were found by different physical motivations, such as \cite{Ein15,LL62,Wein72,Pa48,Tol50,Mo58}.

To understand more about the energy-momentum pseudotensor, analysis of last section is very helpful. One can understand the energy-momentum pseudotensor in terms of the Noether's principle. Starting from the standard Einstein-Hilbert Lagrangian $L_{EH}=\frac{1}{16\pi}R\sqrt{-g}$, one can rediscover Moeller's energy-momentum pseudotensor by calculating the canonical energy-momentum tensor $\theta_{ab}$ \cite{Sz04}. The reson why there exist so many energy-momentum pseudotensor is that there is another ``additional structure" which has not been discussed. That is the boundary term of the Lagrangian. It is well-known that a total divergent term in Lagrangian will not contribute to the dynamical equation, but such term will contribute to the energy-momentum pseudotensor. Adding different boundary term, one can get different energy-momentum pseudotensor by calculating associated canonical energy-momentum tensor $\theta_{ab}$ \cite{CNT15}.

With the help of energy-momentum pseudotensor, one can define the total energy and linear momentum of gravitational field as
\begin{eqnarray}
P^{\mu}(\Sigma)&:=&\int_{\Sigma}(T^{\mu}_{\nu}+t^{\mu}_{\nu})n^{\nu}dx^3\nonumber\\
&=&\frac{1}{16\pi}\int_{\Sigma}\partial_{\alpha}U^{\mu\alpha}_{0}dx^3\nonumber\\
&=&\frac{1}{16\pi}\int_{\partial\Sigma}U^{\mu\alpha}_0dS_{\alpha}.
\end{eqnarray}

The energy-momentum pseudotensor is a quite useful tool for people to study gravitational physics. With help of it, people explained the energy loss of binary system, which was the first indirect evidence for the existence of gravitational wave\cite{bi}. People also used this method got the correct answer for the energy translation of the tidal friction processes, which was gravitational energy being translated into thermal energy caused by the tidal force of planet's elliptic orbits. Such phenomena was first observed on Io, a moon of Jupiter\cite{Tidal79}.

Although the energy-momentum pseudotensor is very useful in many cases, the coordinates dependence is still a disadvantage. To avoid this disadvantage, many research work has been done. One way to solve this problem is to use tetrad instead of metric to rewrite the Lagrangian. It is well-known that the moving frame method is a very powerful method in differential geometry. Compare with the metric $g_{ab}$, the tertad $\{e_{(\alpha)}\}$ has more freedom. Such advantage has been noticed since the beginning of 1960's. In 1961, Moeller began to use ``absolute parallelism" to reconsider pseudotensor problem \cite{Mo61}. Newman, Penrose and their colleagues developed techniques so called ``Newman-Penrose formulism", which was using null tetrad to consider the gravitational radiation \cite{NP62}.

\subsection{Duan's energy-momentum}

Almost the same time, unlike Moeller's absolute parallelism, Prof. Duan used ordinary orthonormal tetrad to study this problem \cite{DZ62}. This is a more natural choice.

Using orthonormal tetrad $\{e_{(\alpha)}\}$, one can rewrite the Einstein-Hilbert Lagrangian as\cite{DZ63}
\begin{eqnarray}
L=\frac{1}{16\pi}\left[(\nabla_{\nu}e^{\nu}_{(\alpha)})(\nabla_{\mu}e^{\mu}_{(\alpha)})
-(\nabla_{\mu}e^{\nu} _{(\alpha)})(\nabla_{\nu}e^{\mu}_{(\alpha)})\right],
\end{eqnarray}
where $e^{\mu}_{(\alpha)}$ is the tetrad, $(\alpha)$ is the tetrad index and $\mu,\nu$ are coordinate index. Based on the idea of Noether theorem, considering the variation of above Lagrangian caused by diffeomophism map, Prof. Duan and his colleagues gave the expression of the total energy and momentum as \cite{DZ63,DW83}:
\begin{eqnarray}
P_{(\alpha)}&=&\int_S\sqrt{g}\ V^{0j}_{(\alpha)}dS_j,\qquad \texttt{greek letter}=0,1,2,3\label{duan}\\
V^{\mu\nu}_{(\alpha)}&=&\frac{1}{8\pi}\left[e^{\mu}_{(\beta)}e^{\nu}_{(\gamma)}\eta_{(\alpha\beta\gamma)}
+(e^{\mu}_{(\alpha)}e^{\nu}_{(\beta)}-e^{\nu}_{(\alpha)}e^{\mu}_{(\beta)})\eta_{(\beta)}\right],\\
\eta_{(\alpha\beta\gamma)}&=&\frac{1}{2}\left\{\frac{\partial e_{\sigma(\alpha)}}{\partial x^{\lambda}}\left[e^{\lambda}_{(\beta)}e^{\sigma}_{(\gamma)}-e^{\lambda}_{(\gamma)}e^{\sigma}_{(\beta)}\right]\right.\nonumber\\
&&\quad +\frac{\partial e_{\sigma(\beta)}}{\partial x^{\lambda}}\left[e^{\lambda}_{(\alpha)}e^{\sigma}_{(\gamma)}-e^{\lambda}_{(\gamma)}e^{\sigma}_{(\alpha)}\right]\nonumber\\
&&\quad \left.+\frac{\partial e_{\sigma(\gamma)}}{\partial x^{\lambda}}\left[e^{\lambda}_{(\beta)}e^{\sigma}_{(\alpha)}-e^{\lambda}_{(\alpha)}e^{\sigma}_{(\beta)}\right]\right\},\\
\eta_{(\alpha)}&=&\eta_{(\beta\alpha\beta)}=\frac{1}{\sqrt{g}}\partial_{\mu}\left[\sqrt{g}e^{\mu}_{(\alpha)}\right],
\end{eqnarray}
where $P_{(0)}$ is the total energy and $P_{(i)}$ are total momentum components.

\subsection{ADM total energy and linear momentum}

Let's remember Weinberg's work\cite{Wein72} again. He tries to introduce some ``Minkwoski coordinates" on space-time such that there is a ``Minkowksi metric" $\eta_{ab}$ as a back ground. If the space-time is asymptotic flat, his idea makes sense. So the asymptotic flat condition also can been as a kind of ``additional structure". For Asymptotic flat space-time $(M,g_{ab})$, if $\Sigma$ is an asymptotic flat Cauchy surface of $(M,g_{ab})$, it should satisfies following conditions :
\bi
\item [(i)]$C$ is a compact subset of $\Sigma$, $\Sigma\backslash C$ is diffeomorphic to $R^3\backslash B_R$ and $\{x^i\}$ is the Euclidean coordinates on this region.

\item[(ii)]In the region $\Sigma\backslash C$, the induced metric $h_{ij}$ and the extrinsic curvature $K_{ij}$ should satisfy following asymptotic conditions:
\begin{eqnarray}
&&h_{ij}=\delta_{ij}+a_{ij},\ a_{ij}\sim(r^{-\alpha}),\ \partial_kh_{ij}\sim O(r^{-\alpha-1}),\ \partial_k\partial_lh_{ij}\sim O(r^{-\alpha-2}),\nonumber\\
&&K_{ij}\sim O(r^{-\alpha-1}),\qquad \partial_kK_{ij}\sim O(r^{-\alpha-2}),\quad \frac{1}{2}<\alpha\le 1\nonumber\\
&&i,j,k,l=1,2,3,\qquad r^2=\sum_{i=1}^3(x^i)^2.  \label{asym}
\end{eqnarray}
\ei
With such asymptotic flat condition, Arnowitt, Deser and Misner\cite{ADM,Liang01,Wald84} introduce the ADM energy and momentum as the total energy and momentum of gravitational field on $\Sigma$
\begin{eqnarray}
&&E_{ADM}:=\frac{1}{16\pi}\int_{S_{\infty}}(\partial_jh_{ij}-\partial_jh_{jj})dS^i,\label{ADME}\\
&&P^{ADM}_i:=\frac{1}{8\pi}\int_{S_{\infty}}(K_{ij}-h_{ij}trK)dS^j.\label{ADMP}
\end{eqnarray}
This is a well accepted definition of the total energy and momentum on $\Sigma$.

\subsection{Bondi's energy-momentum}

It was a serious problem to describe gravitational waves at the early age of general relativity. At the beginning, people just considered the linearized Einstein equation in Minkowski space and found that equation was very similar to Maxwell equation. Analog to electromagnetic wave, people called the wave-like solutions as gravitational waves. But how to describe the concept of gravitational wave in non-linear case was a quite difficult problem. That question was answered by Bondi and his colleagues in 1960'\cite{Bondi62,Sa62}. They gave famous Bondi-Sachs' radiating metrics, which are wave-like, vacuum solutions of the Einstein field equations, to describe the non-linear effect of gravitational waves. Such metrics take the following asymptotical forms
 \begin{eqnarray}
 \begin{aligned}
ds^2_{BS} =&-\big(1-\frac{2M}{r}\big)du ^2 -2du dr+2l du
d\theta +2 \bar l \sin \theta du d\psi \\
 &+r ^2 \Big[(1+\frac{2c}{r})d\theta ^2 +\frac{4d}{r}\sin \theta d
\theta d \psi
+(1-\frac{2c}{r})\sin ^2 \theta d \psi ^2\Big]\\
 &+\mbox{lower order terms},
 \end{aligned}
 \end{eqnarray}
where $u$ is retarded coordinate ($u$-slices are null hypersurfaces), $r=x^1$, $\theta $ and $\psi$ are polar coordinates,
$M$, $c$, $d$ are smooth functions of $u$, $\theta$, $\psi$ defined on $\R \times S ^2$ with regularity condition
$\int _0 ^{2\pi} c(u, \theta, \psi)d\psi =0$ for $\theta =0, \pi$ and all $u$, $l = c _{, \theta} +2c \cot \theta +d _{, \psi} \csc \theta$,
$\bar l = d_{, \theta} +2d \cot \theta -c _{,\psi} \csc \theta$.
With this metric, they also studied how gravitational waves carry away energy and momentum. Denote $n^0=1$, $n^1 =\sin \theta \cos \psi$, $n^2 =\sin \theta \sin \psi$, $n^3=\cos \theta$. At null infinity,
the Bondi energy-momentum on slice $\{u=u_0\}$ are defined as
\begin{eqnarray}
m _\nu (u _0) = \frac{1}{4 \pi} \int _{S ^2} M (u _0, \theta, \psi)n ^{\nu} d S\label{BondiEP}
 \end{eqnarray}
for $\nu =0,1,2,3$. Consider the time derivative of Bondi energy, one has
\begin{eqnarray}
\frac{d}{du} m _{0} (u)=-\frac{1}{4\pi}\int _{S ^2} (c _{,u}) ^2 +(d _{,u})^2 \leq 0.
\end{eqnarray}
This is called Bondi energy-loss formula. The non-increasing property of Bondi energy indicates that the energy can be carried away by gravitational waves and the Bondi energy-momentum can be viewed as the total energy-momentum measured after the loss due to the gravitational
radiation up to that time. Let $|m(u)|=\sqrt{m _1 ^2 (u)+ m _2 ^2 (u)+m _3 ^2 (u)}$. If $|m(u)|\neq 0$, Huang, Yau and Zhang proved the more general energy-loss formula \cite{HYZ}
 \begin{eqnarray}
\frac{d}{du} \Big(m _0 (u) -|m(u)| \Big) = -\frac{1}{4\pi}\int _{S ^2} \Big[(c _{,u}) ^2 +(d _{,u})^2 \Big]
\Big(1- \frac{m_i n^i}{|m|} \Big) \leq 0.
 \end{eqnarray}

It is an interesting question whether the total angular momentum can be detected near or at null infinity. As it is still open to find smooth
Bondi-Sachs' coordinates for Kerr spacetime, it is unclear what rotation means for gravitational systems traveling in the speed of light. In 1962, Newman and Unti introduced another coordinates, which is slightly different to Bondi-Sachs' coordinates, to describe asymptotic flat space-time\cite{NU62}. The Newman-Unti coordinates for Kerr space-time has been worked out in \cite{WB08,Bai07}. Wu and Bai also proved that Kerr metric is the unique stationary, axial-symmetric, asymptotic flat space-time under some algebraic conditions on Weyl curvature\cite{WB08}.

\section{Equivalence}

In this section, we show that the energy-momentum defined by Prof. Duan are the same as the ADM energy-momentum, Bondi energy-momentum . Let's consider the ADM case first. With the asymptotic condition (\ref{asym}), we need to choose a suitable tetrad $e^i_{(\alpha)}$ and a suitable coordinate to do the calculation. For coordinates, we choose the standard ``3+1" coordinates \cite{MTW} as
\begin{eqnarray}
ds^2=-(N^2-|\beta|^2)dt^2+2\beta_idtdx^i+h_{ij}dx^idx^j,
\end{eqnarray}
where
\begin{eqnarray}
&&N=1+f,\quad f\sim O(r^{-1}),\quad\partial_kN\sim O(r^{-1-\varepsilon}),\quad\partial_tN\sim O(r^{-1-\varepsilon}),\nonumber\\
&&\beta_i\sim O(r^{-1}),\quad\partial_k\beta_i\sim O(r^{-1-\varepsilon}), \quad\varepsilon > 0\nonumber\\
&&h_{ij}\ \texttt{is asymptotic  flat}.
\end{eqnarray}
Under such coordinates, one can choose tetrad as
\begin{eqnarray}
&&e^a_{(0)}=n^a=\frac{1}{N}\frac{\partial}{\partial t}-\frac{\beta^i}{N}\frac{\partial}{\partial x^i},\nonumber\\
&&e_a^{(0)}=n_a=-Ndt,\nonumber\\
&&e^{a}_{(k)}=\partial_{k}-\frac{1}{2}a_{kj}\partial_{j}+O(a^2),\nonumber\\
&&e_{a}^{(k)}=dx^{k}+\frac{1}{2}a_{kj}dx^j+\beta^kdt+O(a^2).\label{te}
\end{eqnarray}
With such tetrad and coordinates, one can simplifies Eq .(\ref{duan}) as
\begin{eqnarray}
E&:=&\frac{1}{8\pi}\int_{S_{\infty}}\sqrt{g}\ \left[e^0_{(\beta)}e^j_{(\gamma)}\eta_{(0\beta\gamma)}
+(e^0_{(0)}e^j_{(\beta)}-e^j_{(0)}e^0_{(\beta)})\eta_{(\beta)}\right]dS_j\nonumber\\
&=&\frac{1}{8\pi}\int_{S_{\infty}}N\sqrt{h}\ \left[e^0_{(0)}e^j_{(\gamma)}\eta_{(00\gamma)}
+e^0_{(0)}e^j_{(\beta)}\eta_{(\beta)}-e^j_{(0)}e^0_{(0)}\eta_{(0)}\right]dS_j.
\end{eqnarray}
By definition,
\begin{eqnarray}
\eta_{00\gamma}&=&\frac{1}{2}\left\{\frac{\partial e_{\sigma(0)}}{\partial x^{\lambda}}\left[e^{\lambda}_{(0)}e^{\sigma}_{(\gamma)}-e^{\lambda}_{(\gamma)}e^{\sigma}_{(0)}\right]\right.\nonumber\\
&&\quad +\frac{\partial e_{\sigma(0)}}{\partial x^{\lambda}}\left[e^{\lambda}_{(0)}e^{\sigma}_{(\gamma)}-e^{\lambda}_{(\gamma)}e^{\sigma}_{(0)}\right]\nonumber\\
&&\quad \left.+\frac{\partial e_{\sigma(\gamma)}}{\partial x^{\lambda}}\left[e^{\lambda}_{(0)}e^{\sigma}_{(0)}-e^{\lambda}_{(0)}e^{\sigma}_{(0)}\right]\right\}\nonumber\\
&=&-\frac{\partial e_{\sigma}^{(0)}}{\partial x^{\lambda}}\left[e^{\lambda}_{(0)}e^{\sigma}_{(\gamma)}-e^{\lambda}_{(\gamma)}e^{\sigma}_{(0)}\right].
\end{eqnarray}
Therefore
\begin{eqnarray}
e^0_{(0)}e^j_{(\gamma)}\eta_{(00\gamma)}&=&-e^0_{(0)}e^j_{(\gamma)}\frac{\partial e_{\sigma}^{(0)}}{\partial x^{\lambda}}\left[e^{\lambda}_{(0)}e^{\sigma}_{(\gamma)}-e^{\lambda}_{(\gamma)}e^{\sigma}_{(0)}\right]\nonumber\\
&=&-e^0_{(0)}g^{j\sigma}e^{\lambda}_{(0)}\frac{\partial e_{\sigma}^{(0)}}{\partial x^{\lambda}}
+e^0_{(0)}g^{j\lambda}e^{\sigma}_{(0)}\frac{\partial e_{\sigma}^{(0)}}{\partial x^{\lambda}}\nonumber\\
&=&\partial_j\frac{1}{N}+o(r^{-2}).
\end{eqnarray}
The second term is $e^0_{(0)}e^j_{(\beta)}\eta_{(\beta)}$. Using the asymptotic flat condition of tetrad (\ref{te}), this term becomes
\begin{eqnarray}
e^0_{(0)}e^j_{(\beta)}\eta_{(\beta)}&=&e^0_{(0)}e^j_{(\beta)}\left[e^{\lambda}_{(\beta)}e^{\sigma}_{(\alpha)}-e^{\lambda}_{(\alpha)}e^{\sigma}_{(\beta)}\right]
\frac{\partial e_{\sigma(\alpha)}}{\partial x^{\lambda}}\nonumber\\
&=&e^0_{(0)}g^{j\lambda}e^{\sigma}_{(\alpha)}\frac{\partial e_{\sigma(\alpha)}}{\partial x^{\lambda}}-e^0_{(0)}g^{j\sigma}e^{\lambda}_{(\alpha)}
\frac{\partial e_{\sigma(\alpha)}}{\partial x^{\lambda}}\nonumber\\
&=&\frac{1}{N}e^{\sigma}_{(\alpha)}\frac{\partial e_{\sigma(\alpha)}}{\partial x^{j}}-\frac{1}{N}e^{\lambda}_{(\alpha)}
\frac{\partial e_{j(\alpha)}}{\partial x^{\lambda}}+o(r^{-2})\nonumber\\
&=&\frac{1}{N}e^{\sigma}_{(0)}\frac{\partial e_{\sigma(0)}}{\partial x^{j}}+\frac{1}{N}e^{\sigma}_{(k)}\frac{\partial e_{\sigma(k)}}{\partial x^{j}}\nonumber\\
&&-\frac{1}{N}e^{\lambda}_{(0)}
\frac{\partial e_{j(0)}}{\partial x^{\lambda}}-\frac{1}{N}e^{\lambda}_{(k)}
\frac{\partial e_{j(k)}}{\partial x^{\lambda}}+o(r^{-2})\nonumber\\
&=&-\partial_j\frac{1}{N}+\frac{1}{2N}\partial_jh_{ii}-\frac{1}{2N}\partial_ih_{ij}+o(r^{-2}).
\end{eqnarray}
The third term is $-e^j_{(0)}e^0_{(0)}\eta_{(0)}$, detail calculation shows it only contributes $o(r^{-2})$ terms. Summarize above results and back to the definition of energy (\ref{duan}), it is easy to see that Prof. Duan's definition of $E$ agrees with the ADM energy (\ref{ADME}) for asymptotic flat case.

Now let's consider the total linear momentum $P_{(i)}$. By definition (\ref{duan}),
\begin{eqnarray}
P_{(i)}&=&\int_{S_{\infty}}\sqrt{g}\ V^{0j}_{(i)}dS_j \nonumber\\
&=&\frac{1}{8\pi}\int_{S_{\infty}}\sqrt{g}\left[e^0_{(\beta)}e^j_{(\gamma)}\eta_{(i\beta\gamma)}
+(e^0_{(i)}e^j_{(\beta)}-e^j_{(i)}e^0_{(\beta)})\eta_{(\beta)}\right]dS_j\nonumber\\
&=&\frac{1}{8\pi}\int_{S_{\infty}}N\sqrt{h}\left[\frac{1}{N}e^j_{(0)}\eta_{(i00)}+\frac{1}{N}e^j_{(k)}\eta_{(i0k)}
-\frac{1}{N}e^j_{(i)}\eta_{(0)}\right]dS_j. \label{Pi}
\end{eqnarray}
The value of the first term is vanishes since $\eta_{(i00)}=0$. The second term is
\begin{eqnarray}
\frac{1}{N}e^j_{(k)}\eta_{(i0k)}&=&\frac{1}{N^2}e^j_{(k)}\frac{1}{2}\left[\frac{\partial e^{(i)}_m}{\partial x^0}e^m_{(k)}
-\frac{\partial e^{(i)}_0}{\partial x^l}e^l_{(k)}\right.\nonumber\\
&&\quad \left.+\frac{\partial e^{(k)}_m}{\partial x^0}e^m_{(i)}-\frac{\partial e^{(k)}_0}{\partial x^l}e^l_{(i)}\right]+o(r^{-2}).
\end{eqnarray}
Based on the choice of tetrad (\ref{te}), above equation becomes
\begin{eqnarray}
\frac{1}{N}e^j_{(k)}\eta_{(i0k)}&=&\frac{1}{2N^2}\delta^j_k\left[\frac{1}{2}\partial_0a_{mi}\delta^m_k-\partial_l\beta^i\delta^l_k
+\frac{1}{2}\partial_0a_{mk}\delta^m_i-\partial_l\beta^k\delta^l_i\right]\nonumber\\
&&+o(r^{-2})\nonumber\\
&=&\frac{1}{2}\left[\partial_0h_{ij}-\partial_i\beta_j-\partial_j\beta_i\right]+o(r^{-2}).
\end{eqnarray}
Remember $K_{ij}=\frac{1}{2}L_nh_{ij}$, with the help of asymptotic condition for metric, we get
\begin{eqnarray}
K_{ij}&=&\frac{1}{2}L_nh_{ij}\nonumber\\
&=&\frac{1}{2}\left(\frac{1}{N}\partial_th_{ij}-\frac{\beta^k}{N}\partial_kh_{ij}-h_{ik}\partial_j\frac{\beta^k}{N}-h_{jk}\partial_i\frac{\beta^k}{N}\right)\nonumber\\
&=&\frac{1}{2}\left[\partial_th_{ij}-\partial_i\beta_j-\partial_j\beta_i\right]+o(r^{-2}).
\end{eqnarray}
Then
\begin{eqnarray}
\frac{1}{N}e^j_{(k)}\eta_{(i0k)}&=&K_{ij}+o(r^{-2}).
\end{eqnarray}
Similarly, the third term is
\begin{eqnarray}
-\frac{1}{N}e^j_{(i)}\eta_{0}
&=&-\left[\frac{1}{N}e^j_{(i)}e^{\lambda}_{(0)}e^{\sigma}_{(\alpha)}\frac{\partial e_{\sigma(\alpha)}}{\partial x^{\lambda}}\right]
+\left[\frac{1}{N}e^j_{(i)}e^{\lambda}_{(\alpha)}e^{\sigma}_{(0)}\frac{\partial e_{\sigma(\alpha)}}{\partial x^{\lambda}}\right]\nonumber\\
&=&-\delta^j_i\left[\partial_0N+\frac{1}{2}\delta^m_k\partial_0h_{mk}\right]+\delta^j_i\left[\partial_0N+\partial_k\beta^k\right]+o(r^{-2})\nonumber\\
&=&-\delta^j_itrK+o(r^{-2}).
\end{eqnarray}
Back to Eq.(\ref{Pi}), it is clear that Prof. Duan's total momentum also agrees with the ADM linear momentum in the asymptotic flat case.

The ADM energy characterize the total energy of an asymptotic flat Cauchy surface $\Sigma$. If one wants to consider gravitational radiation, The Bondi energy is need. The Bondi-Sachs metrics are \cite{Bondi62,Sa62}
\begin{eqnarray}
ds^2=(e^{2\beta}\frac{V}{r}-r^2h_{\mu \nu}U^\mu U^\nu )du^2+2e^{2\beta}du dr\nonumber\\
+2r^2h_{\mu \nu}U^\nu du dx^\mu -r^2h_{\mu \nu}dx^\mu dx^\nu, \label{Bondi}
\end{eqnarray}
where $\mu,\nu=2,3,\ U^2=U,\ U^3=W\csc\theta,$
\begin{eqnarray}
(h_{\mu \nu})=\left(
\begin{matrix}
e^{2\gamma}\cosh 2\delta&\sinh2\delta\sin\theta\\
\sinh2\delta\sin\theta&e^{-2\gamma}\cosh2\delta\sin^2\theta
\end{matrix}
\right),
\end{eqnarray}
asymptotic behavior:
\begin{eqnarray}
\gamma &=&\frac{c(u,\theta,\phi)}{r}+(C(u,\theta,\phi)-\frac{1}{6}c^3-\frac{3}{2}cd^2)\frac{1}{r^3}+\cdots,\nonumber\\
\delta &=&\frac{d(u,\theta,\phi)}{r}+(D(u,\theta,\phi)-\frac{1}{6}d^3+\frac{1}{2}c^2d)\frac{1}{r^3}+\cdots,\nonumber\\
\beta &=&-\frac{c^2+d^2}{4r^2}+\cdots,\nonumber\\
W&=&-(\frac{\partial d}{\partial\theta}+2d\cot\theta-\frac{\partial c}{\partial\phi}\csc\theta)\frac{1}{r^2}+\cdots,\nonumber\\
U&=&-(\frac{\partial c}{\partial\theta}+2c\cot\theta+\frac{\partial d}{\partial\phi}\csc\theta)\frac{1}{r^2}+\cdots,\nonumber\\
V&=&r-2M(u,\theta,\phi)+\frac{V_1(u,\theta,\phi)}{r}+\cdots.
\end{eqnarray}
In the Bondi coordinates, one can choose tetrad as \cite{XZ}
\begin{eqnarray}
e^{\mu}_{\ (0)}&=&(1,\frac{1}{2}(e^{-2\beta}-\frac{V}{r}),U,W\csc\theta),\nonumber\\
e^{\mu}_{\ (1)}&=&(-1,\frac{1}{2}(e^{-2\beta}+\frac{V}{r}),-U,-W\csc\theta),\nonumber\\
e^{\mu}_{\ (2)}&=&(0,0,\frac{e^{-\gamma}}{r\sqrt{\cosh{2\delta}}},0),\nonumber\\
e^{\mu}_{\ (3)}&=&(0,0,-\frac{e^{-\gamma}\sinh{2\delta}}{r\sqrt{\cosh{2\delta}}},\frac{e^{\gamma}\sqrt{\cosh{2\delta}}}{r\sin\theta}).
\end{eqnarray}
With the choice of coordinates and tetrad, the definition of energy (\ref{duan}) becomes
\begin{eqnarray}
P_{(0)}&=&\frac{1}{8\pi}\int_{S_{\infty}}\sqrt{g}\left[e^{0}_{(\beta)}e^{j}_{(\gamma)}\eta_{(0\beta\gamma)}
+(e^{0}_{(0)}e^{j}_{(\beta)}-e^{j}_{(0)}e^{0}_{(\beta)})\eta_{(\beta)}\right]dS_j\nonumber\\
&=&\frac{1}{8\pi}\int_{S_{\infty}}\sqrt{g}\left[e^{j}_{(\gamma)}\eta_{(00\gamma)}-e^{j}_{(\gamma)}\eta_{(01\gamma)}
+(e^{j}_{(\beta)}-e^{j}_{(0)}e^{0}_{(\beta)})\eta_{(\beta)}\right]dS_j\nonumber\\
&=&\frac{1}{8\pi}\int_{S_{\infty}}\sqrt{g}\left[e^r_{(1)}\eta_{001}+e^r_{(0)}\eta_{001}
 +e^r_{(0)}\eta_{(0)}+e^r_{(1)}\eta_{(1)}\right]dS.
\end{eqnarray}
By definition, straight calculation will show
\begin{eqnarray}
\eta_{001}&=&(-2M-c\frac{\partial c}{\partial u}-d\frac{\partial d}{\partial u})\frac{1}{r^2}+\mathcal{O}(r^{-3}),\nonumber\\
\eta_0&=&(2+\frac{\partial M}{\partial u})\frac{1}{r}+\bigg{[}]cot\theta\csc\theta\frac{\partial d}{\partial\phi}-\csc^2\theta\frac{\partial^2c}{\partial\phi^2}+2\cot\theta\frac{\partial c}{\partial\theta}\nonumber\\
&&+2\csc\theta\frac{\partial^2d}{\partial\theta\partial\phi}+\frac{\partial^2}{\partial\theta^2}+c(-2\csc^2\theta+\frac{9}{2}\frac{\partial c}{\partial u})-\frac{11}{2}d\frac{\partial d}{\partial u}\nonumber\\
&&+12d\sin\theta\frac{\partial d}{\partial u}-4d\sin^2\theta\frac{\partial d}{\partial u}-\frac{1}{2}\frac{\partial V_1}{\partial u}\bigg{]}\frac{1}{r^2}+\mathcal{O}(r^{-3}),\nonumber\\
\eta_1&=&-\frac{2}{r}-(c\frac{\partial c}{\partial u}+d\frac{\partial d}{\partial u})\frac{1}{r^2}+\mathcal{O}(r^{-3}).
\end{eqnarray}
But the covariant derivatives are used in the definition. Using the asymptotic behaviors of 3-index symbols \cite{Bondi62, Sa62}, we finally obtain
\begin{eqnarray}
P_{(0)}=\frac{1}{4\pi}\lim _{r \rightarrow \infty} \int_{S_{\infty}}\frac{M(u,\theta)}{r^2}dS =\frac{1}{4\pi}\int_{S^2}M(u,\theta) dS.
\end{eqnarray}
So Prof. Duan's energy definition (\ref{duan}) agrees with the Bondi energy at null infinity. Similar straight calculation also can show Prof. Duan's momentum definition also agrees with the Bondi momentum.

\section{Positivity of the total energy}

It was conjectured that the total energy is nonnegative both at spatial infinity and at null infinity for isolated physical systems satisfying the dominant energy condition. For the recent progress of the topics, we refer to \cite{Z15} for details. The positive energy conjecture for the ADM total energy-momentum at spatial infinity was first proved by Schoen and Yau \cite{SY1,SY2,SY3}, later by Witten using spinors \cite{Wi1, PT}. More precisely, they proved

{\em The positive energy theorem: Let $(\Sigma, h, K)$ be asymptotically flat, with possibly a finite number of black holes. Suppose the dominant energy condition holds, then\\
(i) $E \geq \sqrt{P _1 ^2 +P _2 ^2 + P_3 ^2}$ for each end;\\
(ii) That $E=0$ for some end implies $M$ has only one end, and space-time is flat along $\Sigma$.}

The total mass is $\sqrt{E^2 - P _1 ^2 -P _2 ^2 - P_3 ^2}$ which is a Lorentzian invariant. When the positive energy theorem holds, the total mass
is well-defined. In 1999, Zhang proved the positive energy theorem for {\em generalized asymptotically flat initial data set} $(\Sigma, h, p)$ where $p$ is general 2-tensor which is not necessary symmetric \cite{Z1}. Geometrically, the second fundamental form $p$ is nonsymmetric when spacetimes equip with affine connections with torsion. In this case matter translates, meanwhile, it rotates. The idea using connection with torsion was initially duo to E. Cartan \cite{Can, Can1, Can2, Can3}. In \cite{Z1}, Zhang defined the following generalized linear momentum counting both translation and rotation
 \beQ
\bar P _{k} = \frac{1}{8\pi}\int_{S_{\infty}}(p_{ki}-h _{ki} tr_h (p))\ast dx ^i,
\eeQ
and proved the following theorem.

{\em The generalized positive energy theorem: Let $(\Sigma, h, p)$ be generalized asymptotically flat, with a finite number of black holes.
Suppose the generalized dominant energy condition
 \beQ
\frac{1}{2}\big(R +(tr _h (p))^2 -|p|^2\big)\geq\max \big\{|\omega|, |\omega+\chi|\big\},
 \eeQ
holds, $\omega _j = \nabla ^i p _{ji}-\nabla _j tr_h(p)$, $\chi _j =2 \nabla ^i (p _{ij} - p _{ji})$, then\\
(i) $E \geq \sqrt{\bar P_1 ^2 + \bar P _2 ^2 + \bar P _3 ^2}$ for each end;\\
(ii) That $E=0$ for some end implies $M$ has only one end, and
\beQ
 R_{ijkl}+p _{ik}p _{jl}-p _{il}p _{jk} = 0,\quad
 \nabla _i p _{jk} -\nabla _j p _{ik} = 0,\quad
 \nabla ^i (p _{ij} -p _{ji}) = 0.
\eeQ}

In 1974, Regge-Teitelboim defined the total angular momentum for asymptotically flat initial data sets \cite{RT},
 \beQ
 J _k (x_0) =\frac{1}{8\pi}\int_{S_{\infty}} \epsilon _{kuv} (x ^u -x_0 ^u )\pi ^v _{\;i} *dx ^i, \quad
 \pi ^v _{\;i} =K ^{v} _{\;i}-h ^{v} _{\;i} tr_K (h).
 \eeQ
In general, the integrand is O($\frac{1}{r}$) which may not be integrable. This ambiguity resolution requires stronger ``Regge-Teitelboim"
conditions on ends
\beQ
h(x)-h(-x)=O(r^{-3}), \quad \pi (x) +\pi (-x) =O(r^{-3}).
\eeQ
The integrand in Regge-Teitelboim's definition is not tensor, it can not relate the local density to the total angular momentum. To resolve
this difficulty, Zhang defined trace free, non-symmetric tensor of local angular momentum density \cite{Z1}
\beQ
\tilde h ^z _{ij} = \frac{1}{2}\epsilon _{i} ^{\;\;\;uv}\big(\nabla _u \rho _z ^2 \big)\big(K _{vj}- h _{vj} tr _h (K)\big),
\eeQ
where $\rho _z$ is the distance function w.r.t some $z \in M$. If $(\Sigma,g, \tilde h ^z _{ij})$ is generalized asymptotically flat,
Zhang defined the total angular momentum \cite{Z1}
\beQ
J_{k} =\frac{1}{8\pi}\int _{S_{\infty}}\tilde h ^z _{ki}\ast dx ^i.
\eeQ
$J$ is also a geometric quantity which is independent on the choice of coordinates $\{x^i\}$ if $|\nabla \tilde h ^z|$ is integrable.
In Kerr spacetime, $K_{ij}=O(\frac{1}{r^4})$, so the total angular momentum is well-defined and it was found $J=(0, 0, ma)$ \cite{Z1-1}.
By taking $p_{ij}=C \tilde h ^z _{ij}$ for certain constant $C>0$, Zhang proved the Kerr constraint $E \geq C |J|$ under the generalized
dominant energy condition \cite{Z1}. It is worth pointed out that the dominant energy condition does not yield to the Kerr constraint. In \cite{HSW}, Huang, Schoen and Wang showed that it is possible to perturb arbitrary vacuum asymptotically flat initial data sets to new vacuum ones having exactly the same total energy, but with the arbitrary large total angular momentum.

At null infinity, it was conjectured that the Bondi energy is nonnegative. The proofs of this conjecture were claimed by Schoen-Yau using geometric analysis methods, as well as by physicists using Witten's positive energy arguments (see \cite{CJL} and references therein). However, extra conditions are required when two methods are worked out rigorously and completely \cite{HYZ}. In particular, by using the positive energy theorem near null infinity \cite{Z2}, Huang, Yau and Zhang proved

{\em Positivity of Bondi energy: Suppose there exists $u _0$ in vacuum Bondi's radiating spacetime such that $c(u_0)=d(u_0)=0$.\\
(i) $m _{0} (u_0) \geq |m(u_0)|$, and the Bondi energy-momentum loss formula gives $m _{0} (u) \geq |m(u)|$
for all $u \leq u_0$;\\
(ii) If $m _{0} (u_0)=|m(u_0)|$ and there is $u_1 < u_0$ such that $m _{0} (u_1)=|m(u_1)|$, then $c(u_0)=d(u_0)=0$ on $[u_1, u_0]$. Thus the spacetime
is flat in the region $(u_1, u_0]$.}

It ensures that the Bondi mass $\sqrt{m_0 ^2 -m_{1} ^2 -m _{2} ^2 -m_{3} ^2}$ is well-defined.

\section{Positive cosmological constant}

Based on the experimental dates, e.g., Planck 2015\footnote{Planck 2015 results I. Overview of Scientific Results. ArXiv: 1502.01582; XIII.
Cosmological Parameters. ArXiv: 1502.01589.}, the actual universe's metric is asymptotic to the FLRW metric with $k=0$
\beQ
\tilde g _{FLRW} = -dt ^2 + e ^{2Ht} g_\delta , \quad g_\delta=(dx ^1)^2 +(dx ^2)^2+(dx ^3)^2,
\eeQ
where $H>0$ is the Hubble constant, $3H^2 =\rho _m + \Lambda _c$, $\rho_m \cong 0.3156 \times 3H^2$ is the matter density containing dark matter, $\Lambda _c \cong 0.6844 \times 3H^2$ is the real value of cosmological constant representing dark energy. Denote $\Lambda=3H^2$, $\lambda =H^{-1}$.
The initial data set is $(\R ^3, \breve{g}= e ^{2Ht}g_{\delta}, \breve{K}=H \breve{g})$.

To define the total energy-momentum from the Hamiltonian point of view, an asymptotically de Sitter initial data set $(M, g, K)$ should satisfy
\beQ
g-\breve{g}=e ^{2 H t_0} a =O(\frac{1}{r}), \qquad K-\breve{K}=e ^{H t_0} b =O(\frac{1}{r^2})
\eeQ
for constant $t_0$ on ends. Then the ten Killing vectors $U_{\alpha \beta}$ of $\R^{4,1}$ and $a$ and $b$ are used to define the ADM-like total energy-momentum \cite{KT}. But there is no energy-momentum inequalities for them in general. Alternatively, there is different approach to define the total energy-momentum from the initial data set point of view. An initial data set $(M, g, K)$ is {\em $\P$-asymptotically de Sitter} if $g= \P ^2 \bar g$, $h= \P \bar h$ for certain constant $\P>0$, and $(M, \bar g, \bar h)$ is asymptotically flat. Let $\bar E$, $\bar P _{k}$ and $\bar {J _{k}} (z)$ be the total energy, the total linear momentum and the total angular momentum of the end $M_l$ for $(M,\bar g, \bar h)$ respectively. The corresponding quantities for $\P$-asymptotically de Sitter initial data set $(M, g, K)$ are
\beQ
E=\P \bar E,\quad P_{k} =\P^2 \bar P _{k}, \quad J _{k}(z)=\P^2 \bar {J _{k}} (z),
\eeQ
and one of the positive energy theorem proved by Luo, Xie and Zhang \cite{LXZ} is as follows.

{\em The positive energy theorem: Let $(M, g, K)$ be a $\P$-asymptotically de Sitter initial data set in spacetime
${\bf{L}} ^{3,1}$ satisfying the Einstein field equations $${\bf R} _{\alpha \beta}-\frac{{\bf R}}{2}{\bf g} _{\alpha \beta} +\Lambda {\bf g} _{\alpha \beta}=T _{\alpha \beta}.$$ Suppose $tr _g (K) \leq \sqrt{3\Lambda }$. If ${\bf{L}}^{3,1}$ satisfies the dominant energy condition, then\\
(i) $E \geq \sqrt{P_{1} ^2 +P_{2} ^2 +P_{3} ^2}$ for any end;\\
(ii) That $E=0$ for some end implies
\beQ
(M, g, K) \equiv \Big(\R ^3, \P ^2 g_{\delta}, \sqrt{\frac{\Lambda}{3}}\P ^2 g_{\delta} \Big)
\eeQ
and the spacetime ${\bf{L}}^{3,1}$ is de Sitter along $M$.}

It ensures that the total mass $\sqrt{E^2 -P_1 ^2 -P_2 ^2 -P_3 ^2}$ is well-defined.

The two definitions of total energy are the same both in the Hamiltonian formulation and in the initial data set formulation. But the definitions
of the total linear momentum are completely different, while it is finite, defined in terms of $h$, the total linear momentum is infinite in general, defined in terms of $K-\breve{K}$. So, in the Hamiltonian formulation, there should not be any energy-momentum inequality, but $E\geq 0$ if $tr _g (K) \leq \sqrt{3\Lambda }$ in this case.

In 2012, Liang and Zhang constructed counterexamples of the above positive energy theorem while the condition $tr _g (K) \leq \sqrt{3\Lambda }$
is violated \cite{LZ}.

By suitable coordinate transformation, the FLRW metric can be transferred into the retarded coordinates which are used to study the nonlinear theory of gravitational waves. In general, if spacetimes have a family of non-intersecting null hypersurfaces given by the level sets of smooth function $u$, gravitational waves can also be described as the Bondi-Sachs metrics even if the cosmological constant is nonzero. For the positive cosmological constant, $u$ is only continuous with discontinuous derivatives across the cosmological horizon, so the above metrics are valid only inside and outside the cosmological horizon.


The theory of gravitational waves for positive cosmological constant has been studied extensively in recent years. In \cite{GLSWZ}, a detail asymptotic analysis of Bondi-Sachs metrics was provided by assuming Sommerfeld's radiation condition, which is natural in numerical simulations,
\begin{eqnarray}
\begin{aligned}
\gamma&=\frac{c}{r}+\Big(-\frac{1}{6} c^3-\frac{3}{2} d^2 c+C\Big)
   \frac{1}{r^3}+O\Big(\frac{1}{r^4}\Big),\\
\delta&=\frac{d}{r}+\Big(-\frac{1}{6}d^3+\frac{1}{2} c^2d+D\Big)
   \frac{1}{r^3}+O\Big(\frac{1}{r^4}\Big),
\end{aligned}
\end{eqnarray}
with the regularity condition $\int _0 ^{2\pi} c(u, \theta, \psi)d\psi =0$ for $\theta =0, \pi$ and for all $u$. The key point is that, for the positive cosmological constant, $u$ can be only continuous with discontinuous derivatives across the cosmological horizon, so the above expansions are valid only near and inside the cosmological horizon as well as near infinity outside the cosmological horizon.
For the negative cosmological constant, the series expansions are taken near $r=\infty$, the peeling property is not affected even if the
coefficients involve $\Lambda$. But for the positive cosmological constant, the series expansions are taken both near and inside the cosmological horizons where $r \sim \sqrt{3 \Lambda ^{-1}}$ is finite, and near $r=\infty$ outside the cosmological horizon. The peeling property can be affected if the coefficients involve $\Lambda$ near the cosmological horizon. So suitable series expansions are required.

From \cite{GLSWZ, XZ}, we can find the following asymptotic expansions for the vacuum field equations
\begin{eqnarray}
\begin{aligned}
\beta =&B -\frac{c^2+d^2}{4r^2}+O\Big(\frac{1}{r^3}\Big),\\
W=&X+2 e^{2 B} B_{,\phi} \csc \theta\frac{1}{r}+O\Big(\frac{1}{r^2}\Big),\\
U=&Y+2 e^{2 B} B_{,\theta}\frac{1}{r}+O\Big(\frac{1}{r^2}\Big),\\
V=&-\frac{e^{2B}\Lambda}{3}r^3+(\cot\theta Y+\csc \theta X_{,\phi}+Y_{,\theta})r^2\\
  &+e^{2 B} \big(4B_{,\phi}^2 \csc^2\theta+2 B_{,\phi \phi} \csc ^2\theta + 2 B_{,\theta} \cot \theta\\
  &+4 B_{,\theta} ^2 +2 B_{,\theta \theta}+1\big)r-2 M+O\Big(\frac{1}{r}\Big),\\
\end{aligned}
\end{eqnarray}
where $X=\Lambda a \sin \theta $, $Y=\Lambda b \sin \theta $, and $a$, $b$ satisfy
\begin{eqnarray}
\begin{aligned}
a_{,\phi}-\sin \theta b_{,\theta}=-\frac{2}{3} e^{2B} c, \qquad
b_{,\phi}+\sin \theta a_{,\theta}=\frac{2}{3} e^{2B} d.
\end{aligned}
\end{eqnarray}
Thus $a$, $b$ can be determined uniquely by $B$, $c$, $d$ up to some functions $\rho(u)$, $\sigma(u)$.
Moreover, we can choose $B$, $a$, $b$ which are independent on $\Lambda$.

Let $\Psi _k$, $k=0,\dots,4$, be the Newmann-Penrose quantities. It was proved in \cite{XZ} that, under Sommerfeld's radiation condition together with the nontrivial $B$, $X$, $Y$, the following peeling property holds
\begin{eqnarray}
\Psi_k = -\Big[\big(\Psi_{k} ^{5-k}\big)^0 +O\big(\Lambda \big)\Big]\,\frac{1}{r^{5-k}}+O\Big(\frac{1}{r^{6-k}} \Big),
\end{eqnarray}
where coefficients $\big(\Psi_{k} ^{5-k}\big)^0$ are given by $B$, $a$, $b$ and other $\Lambda$-independent functions appeared in series expansions of $\gamma$, $\delta$, $\beta$, $W$, $U$ and $V$. The cosmological constant affects the experimental data only in a scale of $\Lambda$ which can be ignored.

Note that $\big(\Psi_{4} ^{1}\big)^0 \neq 0$ when $B$ is nontrivial, $a=a(u)$, $b=b(u)$ are functions of $u$ which give $c=d=0$. This indicates
that there exist gravitational waves without Bondi news, which may be referred as $B$-gravitational waves. In \cite{XZ}, some nonstationary vacuum Bondi-Sachs metrics are constructed with $\gamma =\delta =0$. Hence $c=d=0$.
\begin{eqnarray}
\begin{aligned}
g=&-\Big[-(C+\cos \theta)^2  \frac{\Lambda}{3}r^2  - \sin ^2\theta\Big(\frac{C'}{C^2 -1}\Big)^2 r^2
-2(C\cos\theta +1)\frac{ C'}{C^2 -1 }r \\
&+C^2 -1 -\frac{2m(C^2 -1)^\frac{3}{2}}{r(C+\cos \theta)}\Big]du^2-2 \big(C+\cos \theta \big)du dr\\
&+2 r \sin \theta \Big(1+\frac{C'}{C^2 -1}r\Big)du d\theta+r^2\Big(d\theta^2 +\sin^2\theta d\phi^2 \Big).
\end{aligned}
\end{eqnarray}
The above metrics have black holes when $m \neq 0$. Moreover, the Newmann-Penrose quantity $\Psi_k$ satisfy
\begin{eqnarray}
\begin{aligned}
\Psi_0&=\Psi_1=0,\qquad \Psi_2=-\frac{m(C^2-1)^{\frac{3}{2}}}{r^3(C+\cos\theta)^3},\\
\Psi_3&=\frac{3m\sin\theta(C^2-1)^{\frac{3}{2}}}{\sqrt{2} r^3(C+\cos\theta)^4},\quad
\Psi_4=-\frac{3m\sin^2\theta(C^2-1)^{\frac{3}{2}}}{r^3(C+\cos\theta)^5}.
\end{aligned}
\end{eqnarray}
They fall faster than usual, which may be missed in the experimental data.

We refer to \cite{HC} for an alternative boundary condition in the axi-symmetric case, without assuming
Sommerfeld's radiation condition, but deforming 2-sphere with
\begin{eqnarray}
\gamma=\Lambda f(u, \theta)+\frac{c(u, \theta)}{r}+O\Big(\frac{1}{r^3}\Big)
\end{eqnarray}
and taking $B=X=Y=0$. The vacuum field equations give
\begin{eqnarray}
f(u, \theta)=f(-\infty, \theta)+\frac{1}{3}\int _{-\infty} ^u c(s, \theta) ds.
\end{eqnarray}
Physically, $f(-\infty, \theta)$ exists. Thus, in order that $f(u, \theta)$ exists for any $u<\infty$ and for $u\rightarrow +\infty$, Bondi news must satisfy $\int _{-\infty} ^u  c(s, \theta) ds <\infty$ for any $u<\infty$ and for $u=+\infty$. This boundary condition is rather restricted which actually excludes gravitational waves with $\int _{-\infty} ^u c(s, \theta) ds =\infty$ or $\int _u ^{+\infty} c(s, \theta) ds =\infty$ for some $u<\infty$ or for $u=+\infty$. In a series of papers \cite{ABK1, ABK2, ABK3, ABK4}, asymptotics with $\Lambda >0$ was discussed in framework of conformal compactification, and the linearization theory as well as the quadrupole formula were derived. Some relevant works on the linearization theory can also be found in \cite{B, DH1, DH2}. The papers \cite{S1, S2} discussed the asymptotic vacuum and the electromagnetism Newman-Penrose equations as well as Bondi mass for
$\Lambda \neq 0$. The boundary condition in \cite{S1, S2} is essentially equivalent to that given in \cite{HC}.

The peeling property for $\Lambda \neq 0$ was proved previously by Penrose in framework of conformal compactification \cite{Penr},
and by Saw without conformal compactification \cite{S1, S2}. It was pointed out in \cite{XZ} that the induced metric on conformal boundary $J^+ $ are
\begin{eqnarray}
\frac{e^{2B} \Lambda}{3}  du^2 +2 Y du d\theta +2X du d\phi +d\theta ^2 +\sin^2 \theta d\phi ^2.
\end{eqnarray}
So the boundary condition given in \cite{GLSWZ, XZ} does not seem to consist with Penrose's framework. This new boundary condition is natural in three aspects that it consists with nontrivial Bondi news, gives rise to the peeling property and features $B$-gravitational waves without Bondi news.

\section{Acknowledgement}
This work is partially supported by the National Science Foundation of China (grants 11571345, 11575286) and the project of mathematics and interdisciplinary
sciences of Chinese Academy of Sciences.


\begin{thebibliography}{GGG}



\bibitem{Sz09}L¡äaszl¡äo B. Szabados,
Living Rev. Relativity 12 (2009) 4.
[Online Article]:
http://www.livingreviews.org/lrr-2009-4

\bibitem{IW94}V. Iyer and R. M. Wald, Phys. Rev. D, 50 (1994) 846.

\bibitem{IW95}V. Iyer and R. M. Wald, Phys. Rev. D, 52 (1995)4430.

\bibitem{Ko59}A. Komar, Phys. Rev. 113(1959)934.

\bibitem{Ein15}A, Einstein, Berlin. Ber. 778 (1915) 154.

\bibitem{LL62}L. D. Landau and E. M. Lifshitz, {\it The Classical Theory of Fields}, 2nd edn. (Addison-
Wesley, Reading, MA, 1962).

\bibitem{Wein72}S. Weinberg, {\it Gravitation and Cosmology}, 1972 (New York: Wiley).

\bibitem{Pa48}A. Papapetrou, Proc. Roy. Irish Acad. (Sect. A) 52A (1948) 11

\bibitem{Tol50}R. C. Tolman, {\it Relativity Thermodynamics and Cosmology}, 1950.

\bibitem{Mo58}C. Moeller, Ann. Phys. (N.Y.) 4 (1958) 347.

\bibitem{Sz04}L. B. Szabados,  Canonical pseudotensors, Sparling¡¯s form and Noether currents, KFKI Report 1991-
29/B, (KFKI Research Institute for Particle and Nuclear Physics (RMKI), Budapest, 1991). Online
version (accessed 29 January 2004):
http://www.rmki.kfki.hu/~lbszab/doc/sparl11.pdf.

\bibitem{CNT15}Chiang-Mei Chen, James M. Nester and Roh-Suan Tung, International Journal of Modern Physics D
Vol. 24, No. 11 (2015) 1530026

\bibitem{bi}R. A. Hulse and J. H. Taylor, Astrophys. J. 195 (1975) L51-L53.

\bibitem{Tidal79}S. J. Peale, P. Cassen and R. T. Reynolds, Science, 203 (1979) 892.

\bibitem{Mo61}C. Moeller,  Mat.-Fys. Skr. K. Danske Vid. Selsk., 1(10) (1961) 1¨C50.

\bibitem{HYZ}W-L. Huang, S.T. Yau, X. Zhang,  Positivity of the Bondi mass in Bondi's radiating spacetimes.
Rend. Lincei Mat. Appl. {\bf 17}, 335-349 (2006).

\bibitem{NP62}E. T. Newman and R. Penrose, J. Math. Phys. 3 (1963) 896.

\bibitem{DZ62}Yi-Shi Duan and Jing-Ye Zhang, ACTA PHYSICA SINCA, 18 (1962) 211.

\bibitem{DZ63}Yi-Shi Duan and Jing-Ye Zhang, ACTA PHYSICA SINCA, 19 (1963) 689.

\bibitem{DW83}Yi-Shi Duan and You-Tang Wang, SCENTIA SINCA (Series A) XXVI (1983) 961.

\bibitem{DF95}Yi-Shi Duan and Shi-Xiang Feng, ACTA PHYSICA SINCA 44 (1995) 1373.

\bibitem{ADM} R. Arnowitt, S. Deser, C. Misner, The dynamics of general relativity. pp.227-264, in {\em Gravitation:
an introduction to current research}, L. Witten, ed. Wiley, New York, 1962.

\bibitem{Liang01} C. B. Liang, {\it Introduction for Differential Geometry and General Relativity}, Beijing Normal University Press, 2001

\bibitem{Wald84} R. M. Wald, {\it General Relativity}, The University of Chicago Press, 1984.

\bibitem{MTW}C. W. Misner, K. S. Thorne and J. A. Wheeler, {\it Gravitation}, W. H. Freeman and Compeny, 1973.

\bibitem{Bondi60}H. Bondi, Nature, 186(1960)535.

\bibitem{Bondi62}H. Bondi, M. G. J. van der Burg and A. W. K. Metzner, Proc. R. Soc. London, Ser. A, 269, 21¨C52 (1962).

\bibitem{Sa62}R. K. Sachs, Waves in asymptotically flat space-time, Proc. R. Soc. London A 270, 103
(1962).

\bibitem{NU62}E. T. Newman and T. W. J. Unti, J. Math. Phys. 3 (1962) 891.

\bibitem{WB08}X. Wu and S. Bai,  Phys. Rev. D78 (2008) 124009.

\bibitem{Bai07}S. Bai, Z. Cao, X. Gong, X. Wu and Y. K. Lau, Phys. Rev. D75 (2007) 044003.




\bibitem{Z15}X. Zhang, International Journal of Modern Physics A, 30, Nos. 28-29 (2015) 1545018 (20 pages)

\bibitem{SY1}R. Schoen, S.T. Yau, On the proof of the positive mass conjecture in general relativity. Commun. Math. Phys. {\bf 65}, 45-76 (1979).

\bibitem{SY2}R. Schoen, S.T. Yau, The energy and the linear momentum of spacetimes in general relativity. Commun. Math. Phys. {\bf 79}, 47-51 (1981).

\bibitem{SY3}R. Schoen, S.T. Yau, Proof of the positive mass theorem. II. Commun. Math. Phys. {\bf 79}, 231-260 (1981).

\bibitem{Wi1}E. Witten, A new proof of the positive energy theorem. Commun. Math. Phys. {\bf 80}, 381-402 (1981).


\bibitem{PT}T. Parker, C. Taubes, On Witten's proof of the positive energy theorem. Commun. Math. Phys. {\bf 84}, 223-238 (1982).

\bibitem{Z1} X. Zhang, Angular momentum and positive mass theorem. Commun. Math. Phys. {\bf 206}, 137-155 (1999).

\bibitem{Can} E. Cartan, Sur une g\'{e}n\'{e}ralisation de la notion de courbure de Riemann et les espaces \`{a} torsion. C. R. Acad. Sci. (Paris)
{\bf 174}, 593-595 (1922).

\bibitem{Can1} E. Cartan, Sur les vari\'{e}t\'{e}s \`{a} connexion affine et la th\'{e}orie de la relativit\'{e} g\'{e}n\'{e}ralis\'{e}e.
Part I: Ann. \'{E}c. Norm. {\bf 40}, 325-412 (1923).

\bibitem{Can2} E. Cartan, Sur les vari\'{e}t\'{e}s \`{a} connexion affine et la th\'{e}orie de la relativit\'{e} g\'{e}n\'{e}ralis\'{e}e.
Part II: Ann. \'{E}c. Norm. {\bf 41}, 1-25 (1924).

\bibitem{Can3} E. Cartan, Sur les vari\'{e}t\'{e}s \`{a} connexion affine et la th\'{e}orie de la relativit\'{e} g\'{e}n\'{e}ralis\'{e}e.
Part III: Ann. \'{E}c. Norm. {\bf 42}, 17-88 (1925).

\bibitem{RT} T. Regge, C. Teitelboim, Role of surface integrals in the Hamiltonian formulation of general relativity. Ann. Phys. {\bf 88}, 286-318 (1974).

\bibitem{Z1-1} X. Zhang, Remarks on the total angular momentum in general relativity. Commun. Theor. Phys. {\bf 39}, 521-524 (2003).

\bibitem{HSW}L-H. Huang, R. Schoen, M-T. Wang, Specifying angular momentum and center of mass for vacuum initial data sets.
Commun. Math. Phys. {\bf 306}, 785-803 (2011).

\bibitem{CJL}P. Chru\'{s}ciel, J. Jezierski, S. Leski, The Trautman-Bondi mass of initial data sets. Adv. Theor. Math. Phys. {\bf 8}, 83 (2004).

\bibitem{Z2}X. Zhang, A definition of total energy-momentua and the positive mass theorem on asymptotically hyperbolic 3-manifolds I.
Commun. Math. Phys. {\bf 249}, 529-548 (2004).

\bibitem{KT} D. Kastor, J. Traschen, A positive energy theorem for asymptotically de Sitter spacetimes. Class. Quantum Gravity
{\bf 19}, 5901-5920 (2002).

\bibitem{LXZ}M. Luo, N. Xie, X. Zhang, Positive mass theorems for asymptotically de Sitter spacetimes. Nucl. Phys. B {\bf 825}, 98-118 (2010).

\bibitem{LZ}Z. Liang, X. Zhang, Spacelike hypersurfaces with negative total energy in de Sitter spacetime. J. Math. Phys. {\bf 53}, 022502 (2012).

\bibitem{GLSWZ} H. Ge, M. Luo, Q. Su, D. Wang, X. Zhang,
Bondi-Sachs metrics and photon rockets, Gen. Relativ. Gravit. {\bf 43}, 2729 (2011).

\bibitem{XZ} F. Xie, X. Zhang, Peeling property of Bondi-Sachs metrics for nonzero cosmological constant, arXiv:1704.06015[gr-qc].


\bibitem{HC} X. He, Z. Cao, New Bondi-type outgoing boundary condition for the Einstein equations with cosmological constant, Int. J. Mod. Phys. D {\bf 24}, 1550081 (2015).

\bibitem{ABK1} A. Ashtekar, B. Bonga, A. Kesavan, Asymptotics with a positive cosmological constant: I. Basic framework, Class. Quantum Grav. {\bf 32}, 025004 (2015).

\bibitem{ABK2} A. Ashtekar, B. Bonga, A. Kesavan, Asymptotics with a positive cosmological constant. II. Linear fields on de Sitter spacetime, Phys. Rev. D {\bf 92}, 044011 (2015).

\bibitem{ABK3} A. Ashtekar, B. Bonga, A. Kesavan, Asymptotics with a positive cosmological constant. III.
The quadrupole formula, Phys. Rev. D {\bf 92}, 104032 (2015).

\bibitem{ABK4} A. Ashtekar, B. Bonga, A. Kesavan, Gravitational Waves from Isolated Systems: Surprising Consequences of a Positive Cosmological Constant, Phys. Rev. Lett. {\bf 116}, 051101 (2016).

\bibitem{B} N. T. Bishop, Gravitational waves in a de Sitter universe, Phys. Rev. D {\bf 93}, 044025 (2016).

\bibitem{DH1} G. Date, S. J. Hoque, Gravitational Waves from Compact Sources in de Sitter Background, Phys. Rev. D {\bf 94},064039 (2016).

\bibitem{DH2} G. Date, S. J. Hoque, Cosmological Horizon and the Quadrupole Formula in de Sitter
Background, arXiv:1612.09511v2 [gr-qc] 5 Apr 2017.

\bibitem{S1} V-L. Saw, Mass-loss of an isolated gravitating system due to energy carried away by gravitational waves with a cosmological constant, Phys. Rev. D {\bf 94}, 104004 (2016).

\bibitem{S2} V-L. Saw, Behaviour of asymptotically electro-$\Lambda $ spacetimes, Phys. Rev. D {\bf 95}, 084038 (2017)

\bibitem{Penr} R. Penrose, Zero rest mass fields including gravitation: asymptotic behavior, Proc. R. Soc. A {\bf 284}, 159 (1965).


\end{thebibliography}
\end{document}